\documentstyle[a4,12pt]{article}
\begin{document}
\setlength{\baselineskip}{20pt}
\setlength{\parskip}{10pt}

\null \ 
\vfill 
{\LARGE Uncertainty principle for proper time and mass II}

\vskip 8mm
{
\setlength{\leftskip}{1cm}
\noindent
Shoju Kudaka \\
{\it Department of Physics, University of the Ryukyus, Okinawa, Japan}\\
Shuichi Matsumoto\footnote{Electronic mail address : shuichi@edu.u-ryukyu.ac.jp} \\
{\it Department of Mathematics, University of the Ryukyus, Okinawa, Japan}
\vfill

------------------------------------------------------------------------------ 

------------------------------------------------------------------------------

}

\vfill@
When we quantize a system consisting of a single particle, the proper time $\tau $ and the rest mass $m$ are usually dealt with as parameters. In the present article, however, we introduce a new quantization rule by which these quantities are regarded as operators in addition to the position and the momentum. Applying this new rule to a scalar particle and to a particle with spin $ 1/2 $, we analyze the time evolution of the operator $\tau $. In the former case, the evolution of the proper time perfectly matches several well-established classical formulae. In the case of the particle with spin $1/2$, our new rule implies that an oscillation appears in the time evolution of the operator $\tau $. This oscillation is similar to Zitterbewegung which is well-known in the ordinary Dirac theory. We formulate one physical effect of this oscillation by considering the interaction with a gravitational field, and estimate how small it is. PACS numbers: 03.65.Bz, 03.20.+i, 04.20.Cv, 04.60.Ds.

\vfill 
\eject

\noindent
{\bf I. INTRODUCTION}

We discussed a clock in a previous article$^1$: Analyzing several processes of measuring the mass of the clock, we concluded that there should exist an uncertainty relation 
\begin{equation}
c^2\Delta m\Delta \tau \approx h \label{eq:uncertainty-re}
\end{equation}
between the rest mass $m$ and the proper time $\tau $ of the clock. 
Moreover, considering the fact that the proper time of a clock is clearly an observable quantity, we suggested that it should be included as a dynamic variable in the appropriate Lagrangian. We showed that, with this assumption, the general momentum conjugate to the proper time $\tau $ is necessarily the rest energy $mc^2$, and that $\tau $ and $m$ should therefore be regarded as operators satisfying the commutation relation
\begin{equation}
[\tau , m]=i\hbar /c^2 \label{eq:tau-m}
\end{equation}
whenever the system is quantized. Of course, the uncertainty relation (\ref{eq:uncertainty-re}) follows from the commutation relation (\ref{eq:tau-m}). 

The principal reasons why we chose to discuss a clock in the previous article are: (i) It seems quite natural that we regard the proper time of a clock as an observable quantity of the system. (ii) Therefore, we can hold, with rather strong conviction, the point of view that the proper time should be dealt with as an operator. (iii) Moreover, some authors$^{2,3}$ have shown that if a system has some function as a clock, then the mass of the system has to fluctuate to an extent.

On the other hand, in the theory of classical mechanics, the concept of the proper time of a particle is introduced through the equation 
$$d\tau ={\sqrt {dt^2-d{\bf x}^2/c^2}},$$ 
where $(t, {\bf x})$ are the coordinates of the particle when observed from an inertial system. Hence it is defined for every particle irrespective of whether the particle has some function as a clock or not. Moreover, it turns out that the analysis of means of measuring the mass of a clock developed in the introduction to Ref. 1 can be applied to every particle, and that the same uncertainty relation (\ref{eq:uncertainty-re}) is valid for every particle, again irrespective of the function of that particle.

From these considerations, we are led to the idea that proper time and  mass should be regarded as operators not only for a clock but also for an arbitrary particle. The objective of the present article is to investigate the validity of this idea. We approach it in the following way: 
\begin{enumerate}
\item We postulate a new quantization rule by which proper time $\tau $ and rest mass $m$ are dealt with as self-adjoint operators that satisfy the commutation relation 
$$[\tau , m]=i\hbar /c^2$$
whenever a system of a single particle is quantized.
\item We apply this new rule to some typical cases and examine whether the implications of the rule match any well established fact. 
\item We must formulate some predicted new phenomena, and discuss detection methods.
\end{enumerate}

Of course, we fully realize that the new rule introduces various problems. For example, it is not trivial that the proper time of such a stable particle as an electron is an observable quantity, and therefore it seems difficult to justify our postulate in such a case. Hence, there is some ambiguity as to the universality of the new rule. 

Furthermore, by using Pauli's reasoning$^{4}$, it can be shown that the mass operator $m$ cannot have any discrete eigenvalue if the commutation relation $[\tau , m]=i\hbar /c^2$ is assumed. This situation seems very unrealistic if we judge from the simpleminded view that mass should be consideded as a discrete constant. On the other hand, as mentioned above, if a system has some function as a clock, then the mass of the system fluctuates to some extent. This suggests that our approach may have some advantage for describing such a system. We do not, however, have any completely satisfactory answer to these problems concerning the mass spectrum at present. 

We hope to discuss these problems in subsequent papers. Here, we focus our attention on the time evolution of the operator $\tau $. (In order to avoid unnecessary confusion, we must emphasize that each Hamiltonian considered in this paper does not include proper time as a variable, and that the Heisenberg representation of the operator $m$ therefore satisfies the equation $dm(t)/dt=0$, that is, the quantity $m$ is conserved.) 

In the next section, we apply our new rule to a system consisting of a scalar particle moving in a gravitational field. As for the time evolution of $\tau $, the implications of the rule completely match both the classical formula for the time delay of a moving particle and the relativistic red-shift formula. In the third section, we apply the rule to a free particle with spin $1/2$. In this case, there appears an oscillation in the time evolution of $\tau $. The oscillation is similar to Zitterbewegung which appears in the orbit of the particle in the ordinary Dirac theory. However, it turns out that we can never observe any physical effect of the oscillation so long as we deal with a free particle. In the fourth section, in order to see some physical effect of this oscillation, we consider a particle with spin $1/2$ moving in a gravitational field. In this situation, we can formulate a physical effect, but, as estimated in the last section, it is very small. We close this paper by showing that proper time cannot be dealt with as an operator in ordinary quantum theories. 

In the present article, we use the system of units in which the equations $c=\hbar =1$ are assumed.           

\vskip 1cm
\noindent
{\bf II. SCALAR PARTICLE IN A GRAVITATIONAL FIELD}

A gravitational field $g_{\mu \nu }$ is assumed to be given, and we consider a scalar particle moving in the field. We assume for simplicity that the field $g_{\mu \nu }$ is so-called static in the following sense: 
\begin{enumerate}
\item The functions $g_{\mu \nu }$ depend only on $x^1, x^2, x^3$.
\item For $j=1, 2, 3$ , we have  $g_{j 0} (=g_{0 j}) =0$.
\end{enumerate}
We consider$^1$ 
\begin{equation}                                       
H=f({\bf x}){\sqrt {m^2+\sum _{j, k=1}^3g^{j k}({\bf x})p_jp_k}}
 \label{eq:Hamiltonian-1}      
\end{equation}
as the Hamiltonian of the system, i.e. the energy of the particle, where $f$ is defined by $g_{0 0}=-f^2 \hskip 2mm (f>0)$ and $t\equiv x^0$ is regarded as the independent variable.

We quantize this system according to our new rule mentioned in the introduction: We consider the proper time $\tau $ and the rest mass $m$ as operators which satisfy the commutation relations  
\begin{equation}
[\tau , m]=i, \hskip 5mm [\tau , x^j]=[\tau , p_j]=[m, x^j]=[m, p_j]=0 \hskip 15mm (j=1, 2, 3),  \label{eq:commutator}
\end{equation}
as well as the position $x^j$ and the momentum $p_j$ satisfying the ordinary commutation relations.

The operators $\tau , m, x^j$ and $p_j\ (j=1, 2, 3)$ can be represented in the Hilbert space of square integrable functions of $\tau, x^1, x^2, x^3$. In particular, the operator $m$ is represented by the differential operator $-i\partial /\partial \tau $, and therefore the rest mass $m$ cannot have any discrete spectrum. As mentioned in the introduction, we do not attempt, in the present article, to discuss the problems concerned with such a mass spectrum. Instead, we focus our attention on the time evolution of the operator $\tau $. We should note here that $m$ is conserved because the Hamiltonian (\ref{eq:Hamiltonian-1}) does not depend on the variable $\tau $.         

For the Heisenberg representation of the operator $\tau $ 
$$\tau (t)=e^{itH}\tau e^{-itH},$$
we find, by using the commutation relations (\ref{eq:commutator}), that \begin{equation}
{\dot \tau }(t)=-ie^{itH}[\tau , H]e^{-itH}=e^{itH}{{mf({\bf x})}\over {\sqrt {m^2+\sum_{j, k=1}^3g^{j k}({\bf x})p_jp_k}}}e^{-itH}, \label{eq:scalar-dot-tau-or}
\end{equation}
where the dot denotes the differential with respect to the parameter $t$.
 
Hence, if the space-time is flat, then we have  
\begin{equation}
{\dot \tau }(t)={m\over {\sqrt {m^2+{\bf p}^2}}}. \label{eq:scalar-dot-tau-fl}
\end{equation}
Consider a state for which the momentum operator ${\bf p}$ and the mass operator $m$ have definite values ${\bf p}'$ and $m'(>0)$ respectively. The mean value of (\ref{eq:scalar-dot-tau-fl}) in such a state is equal to 
\begin{equation}
\langle {\dot \tau }(t)\rangle ={m'\over {\sqrt {m'^2+{\bf p}'^2}}}={\sqrt {1-{\bf v}^2}}, \label{eq:scalar-ex-fl}
\end{equation}
where we have used the expression 
$${\bf p}'={{m'{\bf v}}\over {{\sqrt {1-{\bf v}^2}}}}\hskip 1cm ({\bf v}\equiv {\rm the\ velocity\ of\ the\ particle}).$$
The equation ({\ref{eq:scalar-ex-fl}) completely matches the classical formula for the delay of proper time of a moving particle.  

On the other hand, if the space-time is the Schwarzschild's exterior solution, then the function $f$ has the form$^{5}$  
$$f({\bf x})={\sqrt {1-4a/r}} \hskip 2cm ( a : {\rm a\  constant})$$
when viewed from a suitable coordinate system. Consider a state which satisfies the condition that it is not only well localized around a point but also has a negligible momentum ${\bf p}$ compared with the value of mass $m$. If we consider the mean value of (\ref{eq:scalar-dot-tau-or}) in such a state, then we have 
\begin{equation}                                        
\langle {\dot \tau  }(t)\rangle \approx {\sqrt {1-4a/\langle r\rangle }} \label{eq:scalar-dot-tau-sc}              
\end{equation}
for sufficiently small $t$. The equation (\ref{eq:scalar-dot-tau-sc}) leads to the relativistic red-shift formula.

Thus, in this case, there is no contradiction with such well-established facts concerning the time evolution of proper time.
   
\vskip 1cm
\noindent
{\bf III. FREE PARTICLE WITH SPIN $ 1/2 $}

In this section, we quantize a system of a free particle with spin $1/2$ according to our new rule. We consider the Dirac's Hamiltonian 
\begin{equation}
H=\sum _{j=1}^3\alpha _jp_j+\beta m.\label{eq:Dirac-Hamiltonian}
\end{equation}
(Though the same Hamiltonian is used, the following argument will be different from the ordinary Dirac theory in the sense that mass $m$ is dealt with as an operator.) As is well known, $\alpha _j \ (j=1, 2, 3)$ and $\beta $ are Hermitian matrices such that  
$$[\alpha _j, \alpha _k]_+\equiv \alpha _j\alpha _k+\alpha _k\alpha _j=2\delta _{j k}, \hskip 1cm [\alpha _j, \beta ]_+=0, \hskip 1cm \beta ^2=1.$$
In the present article, we occasionally use the representation  
$$\alpha _j=\pmatrix{ 0 & \sigma _j \cr \sigma _j & 0 \cr }\hskip 5mm (j=1, 2, 3) \hskip 5mm{\rm and}\hskip 5mm \beta =\pmatrix{ I & 0 \cr 0 & -I \cr },$$
where $\sigma _j \ (j=1, 2, 3)$ are the Pauli matrices, and $I$ is the unit matrix of degree 2.

Now, the motion equation for the Heisenberg representation of $\tau $ is
\begin{equation}
{\dot \tau }(t)=-ie^{itH}[\tau , H]e^{-itH}=\beta (t).  \label{eq:spin-dot-tau}
\end{equation}
Furthermore, we have 
\begin{equation}
i{\dot \beta }(t)=[\beta (t), H]=2\beta (t)H-[\beta (t), H]_+=2\beta (t)H-2m   \label{eq:dot-beta-1}
\end{equation}
and  
\begin{equation}
i{\ddot \beta }(t)=2{\dot \beta }(t)H. \label{eq:ddot-beta}
\end{equation}
Solving (\ref{eq:ddot-beta}), we get 
\begin{equation}
{\dot \beta }(t)={\dot \beta }(0)e^{-2itH},   \label{eq:dot-beta-2}
\end{equation}
and combining (\ref{eq:spin-dot-tau}), (\ref{eq:dot-beta-1}) and (\ref{eq:dot-beta-2}), we get 
\begin{equation}
{\dot \tau }(t)=\beta (t)={i\over 2}{\dot \beta }(0)e^{-2itH}H^{-1}+mH^{-1}=\left( \beta -mH^{-1}\right) e^{-2itH}+mH^{-1}, \label{eq:dot-tau-2}
\end{equation}
where $H^{-1}$ denotes the inverse operator of $H$.

The equation (\ref{eq:dot-tau-2}) is similar to the equation 
\begin{equation}
{\dot {\bf x}}(t)={\bf \alpha }(t)={i\over 2}{\dot {\bf \alpha }}(0)e^{-2itH}H^{-1}+{\bf p}H^{-1}=\left( {\bf \alpha }-{\bf p}H^{-1}\right) e^{-2itH}+{\bf p}H^{-1}, \label{eq:dot-x-2}
\end{equation}
which is quite familar in the ordinary Dirac theory$^{6}$. The oscillation which appears in (\ref{eq:dot-x-2}) is named Zitterbewegung.  The fact that a similar oscillation appears also in the time evolution of $\tau $ is one of the results of our new quantization rule.    
(In the ordinary Dirac theory, proper time is considered only as a parameter, and therefore the question of oscillation does not arise. We will argue this point in detail in the last section of this paper.)

The momentum ${\bf p}$, the mass $m$, and the energy $H$ mutually commute, and therefore there exists a state for which those operators ${\bf p}$, $m$, and $H$ have definite values ${\bf p}'$, $m'(>0)$, and $E={\sqrt {m'^2+{\bf p}'^2}}$, respectively. We will take the mean value of $\dot \tau $ in such a state with the positive energy.

The mean value of the last term $m/H$ of (\ref{eq:dot-tau-2}) is equal to
\begin{equation}
\langle {m/H}\rangle ={{m'}\over {\sqrt {m'^2+{\bf p}'^2}}}. \label{eq:last-term}
\end{equation}
On the other hand, the mean value of  
$$\left( \beta -mH^{-1}\right) e^{-2itH}=\beta (t)-mH^{-1}$$
turns out to be $0$. This can be shown in the following way: We denote the projection operator to positive energy states by $P_+$. That is to say, defining an operator $H_+$ by $H_+\equiv {\sqrt {m^2+{\bf p}^2}}$, we set 
$$P_+\equiv {1\over 2}\left( 1 + H{H_+}^{-1}\right) .$$
Then we have 
$$[\beta , \ P_+]={1\over 2}\{ 2\beta H-[\ \beta , \ H\ ]_+\ \} {H_+}^{-1}=\{ \ \beta H-m\ \} {H_+}^{-1},$$
and hence we have 
\begin{equation}
0=P_+[ \ \beta , \ P_+\ ]P_+=P_+\{ \ \beta -mH^{-1}\ \} P_+. \label{eq:vanish}
\end{equation}
This implies that the mean value of the oscillating term $\left( \beta -mH^{-1}\right) e^{-2itH}$ in the positive energy state is equal to $0$.  

Therefore, combining (\ref{eq:last-term}) and the above result, we have
\begin{equation}
\langle {\dot \tau }(t)\rangle ={{m'} \over {\sqrt {m'^2+{\bf p}'^2}}}, \label{eq:free-dot-5}
\end{equation}            
which completely corresponds to the classical formula for the time delay of a moving particle (see, Eq. (\ref{eq:scalar-ex-fl})).

Thus we have shown that although a strange oscillation appears in the time evolution of $\tau $, it does not reveal any physical effect as long as we consider positive energy states.

\vskip 1cm
\noindent
{\bf IV. SPIN $1/2$ PARTICLE IN A GRAVITATIONAL FIELD}

In this section, we consider the interaction with a gravitational field. Then, as shown in the following, we can formulate a new phenomenon induced by the oscillating term. 
  
The covariant Dirac equation in a curved spacetime is given by$^{7}$ 
\begin{equation}
i\gamma ^{\alpha }{V_{\alpha }}^{\mu }\left( \partial _{\mu }+\Gamma _{\mu }\right) -m=0, \label{eq:Dirac-or}
\end{equation}
where $\gamma ^{\alpha }$ are Dirac matrices associated with the spin $1/2$ irreducible representation of the Lorentz group, ${V_{\alpha }}^{\mu }$ is a vierbein, and $\Gamma $ is defined by 
$$\Gamma _{\mu }(x)={1\over 2}\Sigma ^{\alpha  \beta }{V_{\alpha }}^{\nu }(x)\left( \nabla _{\mu }V_{\beta \nu }(x)\right)  \hskip 1cm (\ \Sigma ^{\alpha  \beta }\equiv {1\over 4}[\gamma ^{\alpha }, \gamma ^{\beta }]\ ). $$    
The metric tensor $g_{\mu \nu }$ is related to the Minkowski metric $\eta _{\alpha \beta }={\rm dig}(-1, 1, 1, 1)$ by
$$g_{\mu \nu }(x)={V^{\alpha }}_{\mu }(x){V^{\beta }}_{\nu }(x)\eta _{\alpha \beta }.$$ 
In the present article, we use the representation of the Dirac matrices:
$$\gamma ^0=\beta , \hskip 2cm \gamma ^j=\beta \alpha _j\hskip 1cm  (j=1, 2, 3).$$

We consider the Earth's gravitational field. We can introduce$^8$ a coordinate system $\{ x^{\mu }\} $ in which we approximately have    
$${ds}^2=-\left( 1+2\phi \right) (dx^0)^2+\left( 1-2\phi \right) \sum _{j=1}^3(dx^j)^2 \hskip 1cm (\phi ({\bf x})\equiv gx^3)$$
near the particle, where $g$ denotes the acceleration of gravity. Then the vierbein has the form 
$${V^0}_0=(1+2\phi )^{1/2}, \hskip 5mm {V^j}_k=(1-2\phi )^{1/2}\delta ^j_k\hskip 3mm(j, k=1, 2, 3), \hskip 5mm {\rm the\ \ others}=0.$$
Hence, the Dirac equation (\ref{eq:Dirac-or}) has the form  
$$i\gamma ^0\left( 1+2\phi \right) ^{-1/2}\partial _0+i\left( 1-2\phi \right) ^{-1/2}\sum _{j=1}^3\gamma ^j\partial _j+i\gamma ^{\alpha }{V_{\alpha }}^{\mu }\Gamma _{\mu }-m=0.$$
Defining  $W=-i(1+2\phi )^{1/2}\beta \gamma ^{\alpha }{V_{\alpha }}^{\mu }\Gamma _{\mu }$, we have 
\begin{eqnarray}
i\partial _t&=&-i\left( {{1+2\phi }\over {1-2\phi }}\right)  ^{1/2}\sum _{j=1}^3\alpha _j\partial _j+\left( 1+2\phi \right) ^{1/2}\beta m+W \nonumber \\
&\approx &-i\left( 1+2\phi \right) \sum _{j=1}^3\alpha _j\partial _j+\left( 1+\phi \right) \beta m+W,  \label{eq:Dirac-approx} \\ \nonumber 
\end{eqnarray}
where we have made some approximations assuming that $\phi \ll 1$.

Here we regard the equation (\ref{eq:Dirac-approx}) as the Schr{\"o}dinger equation for the particle, and we consider the right-hand side of (\ref{eq:Dirac-approx}) as the Hamiltonian operator (say, ${\tilde H}$).
The Hamiltonian operator should be Hermitian, so it is natural to think that ${\tilde H}$ has the form  
\begin{eqnarray*}
{\tilde H}&=&{1\over 2}\sum _{j=1}^3{\bf \alpha }_j\left[ \ -i\partial _j, \ 1+2\phi\ \right] _++(1+\phi )\beta m +(W+W^{\dagger })/2\cr
&\equiv &H+\sum _{j=1}^3\alpha _j\left[ \ p_j, \ \phi ({\bf x})\ \right] _++\phi ({\bf x})\beta m+(W+W^{\dagger })/2, \cr
\end{eqnarray*}
where $H$ is the free Hamiltonian given by (\ref{eq:Dirac-Hamiltonian}) and we have set  
$$p_j=-i\partial _j \hskip 1cm  (j=1, 2, 3).$$

Now, for the Heisenberg representation of the operator $\tau $
$$\tau (t)=e^{it{\tilde H}}\tau e^{-it{\tilde H}},$$
we have 
\begin{equation}
{\dot \tau }(t)=-ie^{it{\tilde H}}[\tau , {\tilde H}]e^{-it{\tilde H}}=\beta (t)\{ 1+\phi ({\bf x}(t))\} . \label{eq:tau-dot-curv}
\end{equation}
Let $\psi $ be a state which satisfies the following conditions:
\begin{enumerate}
\item It is an eigenstate of $H$ with a positive eigenvalue (say, $E$).
\item When it is decomposed into the sum of eigenstates of $m$, negative eigenvalues never appear in the spectrum. 
\end{enumerate}
We will take the mean value of ${\dot \tau }$ in such a state $\psi $ at $t=0$. 

First, we have 
\begin{equation}
{\dot \tau }(0)=(\beta -mH^{-1})\{ 1+\phi ({\bf x})\} +mH^{-1}\{ 1+\phi ({\bf x})\}  \label{eq:tau-dot-curv-v1}
\end{equation}
and 
\begin{equation}
\langle mH^{-1}\{ 1+\phi ({\bf x})\} \rangle ={1\over E}\langle m\left( 1+gx^3\right)  \rangle .\label{eq:gra-classical}
\end{equation}
Second, we consider the mean value of $\left( \beta -mH^{-1}\right) \{ 1+\phi ({\bf x})\} $. Note that it is equal to the mean value of $\left( \beta -mH^{-1}\right) \phi ({\bf x})$ by virtue of (\ref{eq:vanish}). We can show by calculation that $\psi $ has the form 
$$ \psi ( \tau , {\bf x})=\pmatrix{ \Psi (\tau , {\bf x}) \cr  {{{\bf \sigma }\cdot {\bf  p}}\over {E+m}}\Psi (\tau , {\bf x}) \cr } $$
when represented as a function of the variables $\tau $ and ${\bf x}$, where $\Psi $ is a two-component function which satisfies the condition 
\begin{equation}
(m^2+{\bf p}^2)\Psi (\tau , {\bf x})=E^2\Psi (\tau , {\bf x}). \label{eq:condition}
\end{equation}
Using this form, we have 
\begin{eqnarray*}
&&\langle \psi \vert \left( \beta -mH^{-1}\right) \phi ({\bf x})\vert \psi \rangle \cr
&&=\langle \psi \vert \left( \beta -m/E\right) \phi ({\bf x})\vert \psi \rangle \cr
&&=g\int d\tau d{\bf x}\Psi ^{\dagger }(\tau , {\bf x})\left[ \left( 1-{m\over E}\right) x^3-{1\over {E(E+m)}}({\bf \sigma }\cdot {\bf p})x^3({\bf \sigma }\cdot {\bf p})\right] \Psi (\tau , {\bf x})\cr
&&=g\int d\tau d{\bf x}\Psi ^{\dagger }(\tau , {\bf x})\left[ \left( 1-{m\over E}\right) x^3-{{({\bf \sigma }\cdot {\bf p})^2}\over {E(E+m)}}x^3-{{{\bf \sigma }\cdot {\bf p}}\over {E(E+m)}}[x^3, \ {\bf \sigma }\cdot {\bf p}]\right] \Psi (\tau , {\bf x})\cr
&&=g\int d\tau d{\bf x}\Psi ^{\dagger }(\tau , {\bf x})\left[ \left( 1-{m\over E}\right) x^3-{{{\bf p}^2}\over {E(E+m)}}x^3-i{{{\bf \sigma }\cdot {\bf p}}\over {E(E+m)}}\sigma _3\right] \Psi (\tau , {\bf x}).\cr
\end{eqnarray*}
The first term in the last line offsets the second one by virtue of the condition (\ref{eq:condition}), and the product $i(\sigma \cdot {\bf p})\sigma _3$ in the third term can be replaced by 
$$i[\sigma \cdot {\bf p}, \ \sigma _3]/2=-({\bf \sigma }\times {\bf p})_3$$
because the mean value should be a real number. Eventually we have 
\begin{equation}
\langle \psi \vert \left( \beta -mH^{-1}\right) \phi ({\bf x})\vert \psi \rangle =g\int d\tau d{\bf x}\Psi ^{\dagger }(\tau , {\bf x}){{({\bf \sigma }\times {\bf p})_3}\over {E(E+m)}} \Psi (\tau , {\bf x}). \label{eq:Zitt-mean}
\end{equation}
Combining (\ref{eq:gra-classical}) and (\ref{eq:Zitt-mean}), we finally have 
\begin{equation}
\langle {\dot \tau }(0)\rangle =g\int d\tau d{\bf x}\Psi ^{\dagger }(\tau , {\bf x}){{({\bf \sigma }\times {\bf p})_3}\over {E(E+m)}} \Psi (\tau , {\bf x})+{1\over E}\langle m(1+gx^3)\rangle . \label{eq:final-mean}
\end{equation}

Let us assume that the operators ${\bf p}$ and $m$ have very sharp values ${\bf p}'$ and $m'(>0)$, respectively, in the state $\psi $. Then, by virtue of (\ref{eq:condition}), $E$ can be estimated as 
$$E\approx {\sqrt {m'^2+{\bf p}'^2}},$$
and the equation (\ref{eq:final-mean}) can be rewritten as 
\begin{equation}
\langle {\dot \tau }(0)\rangle \approx 
{g\over {E(E+m')}}\int d\tau d{\bf x}\Psi ^{\dagger }
(\tau , {\bf x})({\bf \sigma }\times {\bf p}')_3\Psi (\tau , {\bf x})+{{m'}\over E}(1+g\langle x^3\rangle ). \label{eq:final-mean-approx}
\end{equation}
The second term on the right-hand side of (\ref{eq:final-mean-approx}) is estimated as 
$${{m'}\over E}(1+g\langle x^3\rangle )\approx {{m'}\over {\sqrt {m'^2+{\bf p}'^2}}}(1+g\langle x^3\rangle ),$$
and it corresponds to the right-hand side of  (\ref{eq:free-dot-5}). On the other hand, the first term is one effect of the oscillating term appearing in (\ref{eq:dot-tau-2}). 

Thus we have succeeded in formulating a new phenomenon induced by the oscillating term in (\ref{eq:dot-tau-2}) by considering the interaction with a gravitational field. 

We here have to note that the normalization condition $\Vert \psi \Vert =1$ can be expressed in the form 
\begin{equation}
\int d\tau d{\bf x}\Psi ^{\dagger }(\tau , {\bf x}){{2E}\over {E+m}}\Psi (\tau , {\bf x})=1. \label{eq:nomalized-condition}
\end{equation}

\vskip 1cm
\noindent
{\bf V. DISCUSSIONS}

Now we would like to estimate how small the first term on the right-hand side of (\ref{eq:final-mean-approx}) is. We consider how small it is compared to the second term:
\begin{eqnarray}
&&\hskip -1cm {g\over {E(E+m')}}\int d\tau d{\bf x}
\Psi ^{\dagger }({\bf \sigma }\times {\bf p}')_3\Psi 
\times \left( {{m'}\over E}\right) ^{-1} \nonumber \\
&=&{g\over {m'}}{1\over {E+m'}}\int d\tau d{\bf x}\Psi ^{\dagger }({\bf \sigma }\times {\bf p}')_3\Psi \label{eq:our-ratio} \\
&\leq &{g\over {m'}}{{\vert {\bf p}'\vert }\over E}
\leq {g\over {m'}}, \nonumber \\ \nonumber 
\end{eqnarray}
where we have used the condition (\ref{eq:nomalized-condition}). 

For an electron and a muon, we have 
$${g\over {m_e}}={{g\hbar }\over {m_ec^3}}\approx 4\times 10^{-29} \hskip 1cm {\rm and}\hskip 1cm {g\over {m_{\mu }}}={{g\hbar }\over {m_{\mu }c^3}}\approx 2\times 10^{-31},$$
respectively, which are rather small. Even if we consider the gravitational acceleration on the surface of the sun (say, $g_s$), the order of these values hardly changes because $g_s/g\approx 10$. 

The expressions
$$g_{0 0}\approx -\left( 1+2\phi \right) \hskip 1cm {\rm and}\hskip 1cm g_{j j}\approx \left( 1-2\phi \right) \hskip 5mm (j=1, 2, 3) $$
mean that $g$ can be interpreted as the rate of change of the metric tensor per unit length. Therefore we can suppose that the ratio 
(\ref{eq:our-ratio}) can be the order of $1$ only when the space-time is so curved that the metric tensor varies conspicuously even in the region of the Compton wavelength $1/m'$ of the particle.

The first term on the right-hand side of (\ref{eq:final-mean-approx}) has the remarkable characteristic that it depends on the directions of the spin and the momentum. We might be able to devise some good experiment to observe this effect by making use of this characteristic. In any case, it is certain that we have to devise a rather precise method in order to see such a small effect. Unfortunately, at present, the authors cannot suggest such a suitable experiment, but sincerely hope that experimentalists will pay attention to this subject. 

Thus we have discussed a new quantization rule that treats the proper time and the rest mass as operators. The rule led us to the existence of an oscillating term in the time evolution of the proper time of a spin $1/2$ particle. We have formulated one physical effect of this term, and have estimated  how small it is. 

Before closing this paper, in order to emphasize the significance of our argument, we need to confirm that those quantities cannot be dealt with as operators in ordinary quantum theories. To make our discussion as concrete as possible, we examine whether the proper time can be dealt with as an operator in the ordinary Dirac theory.

In the classical relativistic mechanics, the proper time $\tau $ is defined by the equation 
\begin{equation}
{\dot \tau }(t)={\sqrt {1-{{\dot {\bf x}}(t)}^2}}, \label{eq:dot-tau-dis}
\end{equation}
where ${\bf x}(t)$ represents the orbit of the particle. If we attempted to deal with the proper time as an operator in the Dirac theory, we would first think of utilizing this definition, that is, we would try to define the \lq \lq time evolution of operator $\tau $'' by replacing ${\dot {\bf x}}$ in (\ref{eq:dot-tau-dis}) by the operator ${\dot {\bf x}}(t)=\alpha (t)$ given by (\ref{eq:dot-x-2}). However, this method can be dismissed trivially, because the equation (\ref{eq:dot-x-2}) tells us that 
$${{\dot {\bf x}}(t)}^2={{\bf \alpha }(t)}^2=3, $$
and therefore the above mentioned substitution only leads us to the meaningless equation
$${\dot \tau }(t)={\sqrt {1-{{\bf \alpha }(t)}^2}}={\sqrt {-2}}.$$
If we must replace ${\dot {\bf x}}$ in (\ref{eq:dot-tau-dis}) by some operator, it probably should be ${\bf p}H^{-1}$, which is only a part of the operator ${\dot {\bf x}}(t)$. In fact, this substitution leads us to 
$${\dot \tau }(t)={\sqrt {1-{\bf p}^2/H^2}}={m\over {\sqrt {m^2+{\bf p}^2}}}.$$
The evolution thus defined corresponds to the classical formula for the time delay of the moving particle, but is not desirable in the sense that it is not equivalent to the time evolution (\ref{eq:dot-x-2}) of the operator ${\bf x}$. 
That is to say, even if we attempt to utilize (\ref{eq:dot-tau-dis}) in order to deal with the proper time as an operator, the approach does not work well. 

We have been unable to find any desirable means of representing proper time as an operator, despite a reexamination of both the theory of relativity and the ordinary Dirac theory. We must conclude therefore that it only has meaning in the latter theory as a parameter.

\vskip 1cm 
\noindent
$^1$ S.Kudaka and S.Matsumoto, \lq \lq Uncertainty principle for proper time and mass'' to appear in J. Math. Phys. (1999). \hfill \break
$^2$ H.Salecker and E.P.Wigner, Phys. Rev. {\bf 109}, 571 (1958). \hfill \break
$^3$ R.Penrose, Gen. Relativ. Gravit. {\bf 28}, 581 (1996). \hfill \break 
$^4$ W.Pauli, in {\it Handbuch der Physik}, 2nd edition, Vol.{\bf 24}  (Springer, Berlin, 1933), pp.83-272; M.Jammer, {\it The Philosophy of Quantum Mechanics} (John Wiley \& Sons, Inc., 1974).\hfill \break
$^5$ K.Schwarzschild, S.B. preuss. Akad. Wiss. 189 (1916). \hfill \break
$^6$ A.Messiah, {\it M{\'e}canique Quantique} (Dunod, Paris, 1959). \hfill \break
$^7$ N.D.Birrell and P.C.W.Davies, {\it Quantum fields in curved space} (Cambridge University Press, Cambridge, 1982).\hfill \break
$^8$ C.W.Misner, K.S.Thorne, and J.A.Wheeler, {\it Gravitation} (W.H.Freeman and Company, New York, 1973). 

\vfill
\end{document}